\documentclass[a4paper,oneside,final,notitlepage,onecolumn,11pt]{article}

\usepackage{amsmath,amssymb}

\begin{document}

\title{Hyperbolic capture of compact binaries}

\author{M\'{a}ty\'{a}s Vas\'{u}th\thanks{email: vasuth.matyas@wigner.mta.hu}
\\
\small WIGNER RCP, RMKI \\
\small  H-1121 Budapest, Konkoly Thege Mikl\'os \'ut 29-33.\\
\small Hungary}
\date{\vspace{-5ex}}
\maketitle

\begin{abstract}
Hyperbolic encounters of compact objects are common interactions in
dense environments. During this process a significant amount of
gravitational radiation is emitted depending on the parameters of
the system. Here we give a parametric description of the radial 
motion valid for general binary orbits and the radiative energy and 
angular momentum losses for binaries with spinning components. 
\end{abstract}

\maketitle

\section{Introduction}

Coalescing compact binaries with high orbital eccentricity are among the promising sources of gravitational waves. With the completion of the advanced gravitational wave detectors the detection of these sources are expected within the next few years.

The orbital evolution of inspiralling binaries can be conveniently described by the post-Newtonian (PN) approximation \cite{BlanchetLRR} while the gravitational potential and source velocities can be considered as small parameters. In this regime the motion of the binary is well approximated by perturbed Keplerian orbits. The leading order contributions to the evolution of binaries on open orbits are well described. The energy flux, the total energy emitted in GWs and the quadrupole contribution to the waveform during the hyperbolic interaction is given e.g. in \cite{Capociello}. These results were generalized for all the three types of binary orbits in \cite{Capociello2}. Moreover, the authors have estimated the rate of hyperbolic encounters in globular clusters and the Galactic Center. Extending \cite{Capociello} the authors of \cite{VJK} have given the the leading order energy flux and the total energy radiated during the hyperbolic interaction, and calculated the radiated energy spectrum for this type of encounters.

In \cite{Turner} gravitational radiation from a system of two point-masses in unbound orbits (orbital eccentricity e $\sim$ 1) is analyzed. Waveforms of the multipole amplitudes are presented for bound and unbound orbits, moreover, the emitted energy and the energy spectrum are also presented.

In this paper we extend the description of the binary system up to the 1.5PN order, with the inclusion of the 1PN and spin-orbit (SO) contributions. The conserved quantities characterizing the orbital evolution is described in Sec. 2, where, based on the radial motion, a suitable parametrization valid for all three types of binary orbits is introduced. In Sec. 3 we give the radiative change of the energy and angular momentum in terms of the conserved quantities. 

Throughout the paper we use unit in which $G=c=1$.

\section{Orbital evolution and the parametrization of the orbit}

In the post-Newtonian approximation the motion of the binary is governed by a generalized Lagrangian. For a collection of the relativistic linear contributions to the leading order Keplerian dynamics see e.g. \cite{Kepler}. In our approximation we have considered only the 1PN and SO terms. Their specific expressions can be found e.g. in \cite{DDAnnInst} and \cite{KWW93}, respectively. In the SO part we are applying the covariant spin supplementary condition \cite{BOC74,Kidder}.

The constants of motion, derived from the generalized Lagrangian, are the energy $E$ of the system \cite{DDAnnInst,GPV3},
\begin{eqnarray}\label{energy}
E&=&\frac{\mu v^{2}}{2}-\frac{m\mu }{r}+E_{PN}+E_{SO}\ ,  \nonumber \\
E_{PN}&=&\frac{3}{8}\left( 1-3\eta \right) \mu v^{4}+\frac{m\mu }{%
2r}\left[ \left( 3+\eta \right) v^{2}+\eta \dot{r}^{2}+\frac{m}{r}\right] \ ,  \notag \\
E_{SO}&=& \frac{\mathbf{L}\cdot{\mbox{\boldmath $\sigma$}}}{r^{3}}\ .
\end{eqnarray}
and the total angular momentum $\mathbf{J}=\mathbf{L}+\mathbf{S}$. Here $m=m_1+m_2$, $\mu=m_1m_2/m$, $\eta=\mu/m$, $r=\vert\mathbf{r}\vert $, $\mathbf{v}=\dot{\mathbf{r}}$, $v=\vert\mathbf{v}\vert $,  and
\begin{eqnarray}
\mathbf{S} = \mathbf{S}_1 + \mathbf{S}_2\ ,\quad {\mbox{\boldmath $\sigma$}}
= \frac{m_2}{m_1}\mathbf{S}_1 + \frac{m_1}{m_2}\mathbf{S}_2 \ .
\end{eqnarray}
The orbital angular momentum $\mathbf{L}=\mathbf{L}_N+\mathbf{L}_{PN}+\mathbf{L}_{SO}$,
\begin{eqnarray}  \label{angmom}
\mathbf{L}_{N} &=& \mu \mathbf{r}\times{\mathbf{v}} \ ,  \notag \\
\mathbf{L}_{PN} &=& \left[\left( 1-3\eta \right)\frac{v^{2}}{2} +
\left( 3+\eta \right)\frac{m }{r} \right]\mathbf{L}_{N} \ ,  \notag \\
\mathbf{L}_{SO} &=& \frac{\mu }{m}\left\{\frac{m}{r^3} \mathbf{r}\times\left[\mathbf{r}\times (2\mathbf{S}+{\mbox{\boldmath $\sigma$}})\right] -\frac{1}{2}\mathbf{v}\times (\mathbf{v}\times {\mbox{\boldmath $\sigma$}}) \right\} \ ,
\end{eqnarray}
undergoes a pure rotation \cite{GPV3} and its magnitude
$L^2=L^2_N+L^2_{PN}+L^2_{SO}$ is contant,
\begin{eqnarray}  \label{Lsquare}
L^2_{N} &=& \mu^2r^2\left(v^2 - \dot{r}^2 \right) \ ,  \notag \\
L^2_{PN} &=& 2L^2_{N}\left[\left( 1-3\eta \right)\frac{v^{2}}{2} +
\left( 3+\eta \right)\frac{m }{r} \right] \ ,  \notag \\
L^2_{SO} &=& \frac{2E}{m}\mathbf{L}\cdot{\mbox{\boldmath $\sigma$}} - \frac{4\mu}{r}\mathbf{L}\cdot\mathbf{S} \ .
\end{eqnarray}
Note that in the SO part of Eqs. (\ref{energy}) and (\ref{Lsquare}) we have replaced the Newtonian orbital angular momentum $\mathbf{L}_N$ with $\mathbf{L}$ by introducing higher order terms neglected in our approximation.

In agreement with \cite{DDAnnInst,Kepler,GPV3,MFV}, Eqs. (\ref{energy}) and (\ref{Lsquare}) can be inverted in order to express the dynamical quantities $v^2$,
\begin{eqnarray}\label{v2}
v^{2} &=&\frac{2E}{\mu }+\frac{2m}{r} +v_{PN}^{2}+v_{SO}^{2}\ ,  \notag  \label{vsquare} \\
v_{PN}^{2} &=&3(3\eta-1)\frac{E^{2}}{\mu ^{2}}+2(7\eta-6)\frac{Em}{\mu r}+5(\eta-2) \frac{m^{2}}{r^{2}}
+ \frac{\eta m L^2}{\mu^2 r^3} \ ,  \notag \\
v_{SO}^{2} &=& -\frac{2}{\mu r^{3}}\mathbf{L}\cdot{\mbox{\boldmath $\sigma$}}\ ,
\end{eqnarray}
and $\dot{r}^2$ in terms of the conserved quantities, resulting in the following expression
\begin{eqnarray}\label{radial}
\dot{r}^{2} &=& \frac{2E}{\mu }+\frac{2m}{r}-\frac{L^{2}}{\mu ^{2}r^{2}} + \dot{r}_{PN}^{2} + \dot{r}_{SO}^{2}\ ,  \notag \\
\dot{r}_{PN}^{2} &=& 3(3\eta -1)\frac{E^{2}}{\mu ^{2}}+2(7\eta
-6)\frac{Em}{\mu r}+2(1-3\eta)\frac{EL^{2}}{\mu ^{3}r^{2}}+5(\eta -2)\frac{%
m^{2}}{r^{2}}+(8-3\eta)\frac{mL^{2}}{\mu ^{2}r^{3}}\ ,  \notag \\
\dot{r}_{SO}^{2} &=& \frac{2}{m\mu r^2}{\bf L\cdot} \left[\frac{E}{\mu}{\mbox {\boldmath $\sigma$}} - \frac{m}{r}\left(2{\bf S}+{\mbox {\boldmath $\sigma$}} \right)\right]\ .
\end{eqnarray}

The turning points of the radial motion are defined by the relation $\dot{r}^{2}=0$. The solution, up to the required order, can be written as \cite{Kepler,MFV}
\begin{eqnarray}\label{rmaxmin}
r_{_{min}^{max}} &=&\frac{m\mu \pm A_{0}}{-2E}+\delta r_{_{min}^{max}}^{PN}++\delta r_{_{min}^{max}}^{SO}\
,  \notag \\
\delta r_{_{min}^{max}}^{PN} &=&(\eta -7)\frac{m}{4}\pm (\eta +9)\frac{%
m^{2}\mu }{8A_{0}}\mp (3\eta -1)\frac{A_{0}}{8\mu }\ , \notag \\
\delta r_{_{min}^{max}}^{SO} &=& \pm\frac{E}{m\mu A_0}{\bf L\cdot}{\mbox{\boldmath $\sigma$}}
 + \frac{\mu\left(A_0\mp m\mu\right) }{L^2 A_{0}}{\bf L\cdot}\left(2{\bf S}+{\mbox {\boldmath $\sigma$}} \right)
\ .
\end{eqnarray}%
Here we have introduced $A_{0}^{2}=m^{2}\mu ^{2}+2EL^{2}/\mu $, which, in leading order, coincides with the magnitude of the Laplace-Runge-Lenz vector.

With the use of the conchoidal transformation, see e.g. \cite{DDAnnInst}, the solution of the radial equation Eq. (\ref{radial}) is given in terms of the parameter $\chi$,
\begin{eqnarray}\label{rchi}
r &=& \frac{L}{\mu (m\mu+A_0\cos\chi) } + \frac{r_{PN}}{2\mu^3A_0 (m\mu+A_0\cos\chi)^2 } + \frac{r_{SO}}{m\mu^2L^2A_0 (m\mu+A_0\cos\chi)^2 }\ ,  \notag \\
r_{PN} &=& m\mu^2A_0\left[2EL^2(4-\eta)-m^2\mu^3(4+\eta)\right]  \notag \\
&&+\left[{E}^2 L^4 (3 \eta -1)+2 {E} L^2 m^2 \mu ^3  (3 \eta -10)
+ 2 m^4 \mu ^6 (\eta -6) \right]\cos\chi \notag \\
&&- m\mu^3A_0^3(8-3\eta)\cos^2\chi\ , \notag \\
r_{SO} &=& m\mu^2A_0\left[3m^2\mu^3{\bf L\cdot}\left(2{\bf S}+{\mbox {\boldmath $\sigma$}} \right)-2EL^2{\bf L\cdot}{\mbox{\boldmath $\sigma$}}\right]
\notag \\
&&-\left[2{E}^2 L^4 {\bf L\cdot}{\mbox{\boldmath $\sigma$}}
-4 {E} L^2 m^2 \mu ^3  {\bf L\cdot}(3{\bf S}+{\mbox {\boldmath $\sigma$}})
- 4 m^4 \mu ^6 {\bf L\cdot}(2{\bf S}+{\mbox {\boldmath $\sigma$}}) \right]\cos\chi \notag \\
&&+ m\mu^3A_0^3{\bf L\cdot}(2{\bf S}+{\mbox {\boldmath $\sigma$}})\cos^2\chi\
\ .
\end{eqnarray}
The parameter $\chi$ is related to the coordinate time $t$ as
\begin{eqnarray}\label{tchi}
\frac{dt}{d\chi} &=& \frac{\mu r^2}{L}\left\{1+\frac{1}{\mu L^2}\frac{dt}{d\chi}_{\vert PN}
+\frac{1}{m L^4}\frac{dt}{d\chi}_{\vert SO} \right\} ,  \notag \\
\frac{dt}{d\chi}_{\vert PN} &=& EL^2(1-3\eta)-m^2\mu^3(1-\eta)\ ,  \notag \\
\frac{dt}{d\chi}_{\vert SO} &=& EL^2{\bf L\cdot}{\mbox{\boldmath $\sigma$}}
- m^2\mu^3{\bf L\cdot}(2{\bf S}+{\mbox {\boldmath $\sigma$}}) \ .
\end{eqnarray}
We note that, independently of the value of the energy $E$, the above parametric solution is valid for all three type of binary orbits. The boundary of the orbit is defined by the relation $\cos\chi_0=min(-1,-m\mu/A_0)$.

\section{Radiative change of the energy and angular momentum}
The orbital dynamics of the binary is characterized by the conserved quantities of the motion. Due to radiation reaction these conserved quantities will change in an adiabatic treatment on a timescale much larger than the orbital period.

The energy and the angular momentum losses are expressed in terms of symmetric trace-free radiative multipole moments, see e.g. \cite{BlanchetLRR}. For point particles their expressions are given in \cite{Kidder} and, for completeness, summarized in the Appendix.

Evaluating the expressions of the energy and the angular momentum losses, Eqs. (\ref{Erad},\ref{jrad}), with use of (\ref{v2}) and (\ref{radial}) and neglecting higher-order terms above 1.5PN order the total radiated energy is
\begin{eqnarray}
{\frac{dE}{dt}} &=& - {\frac{8m^{2}}{15r^{6}}}\left
(2E\mu r^{2}+2m\mu^{2}r+11L^{2} \right ) + \frac{dE_{PN}}{dt} + \frac{dE_{SO}}{dt}\ , \nonumber \\
\frac{dE_{PN}}{dt}  &=& {\frac{2m^{2}}{105\mu^2r^{8}}}\Bigl\{
4E^2\mu^2(149-135\eta)r^{4} + 8E m\mu^3(116-131\eta)r^{3} \nonumber \\
&&+4\mu\left[2EL^2(364-465\eta) + m^2\mu^3(86-111\eta)\right]r^2 \nonumber \\
&&+ 4L^2 m\mu^2(1924-813\eta)r - 3L^4(687-620\eta) \Bigr\} \nonumber \\
\frac{dE_{SO}}{dt}  &=& {{\frac{8m}{15\mu r^{8}}}{\bf L\cdot}\left[
2E\mu r^{2}(10{\bf S}+3{\mbox {\boldmath $\sigma$}}) - 6m\mu^{2}r(2{\bf S}+3{\mbox {\boldmath $\sigma$}})
+3L^{2}(9{\bf S}+17{\mbox {\boldmath $\sigma$}}) \right ]} \ .
\end{eqnarray}

The radiative change of $L$ can be evaluated by $LdL/dt={\bf L}\cdot d{\bf L}/dt$ and it
has the form
\begin{eqnarray}
{\frac{dL}{dt}} &=& {{\frac{8mL}{5\mu r^{5}}}\left
(2E\mu r^{2}-3L^{2}\right )} + \frac{dL_{PN}}{dt} + \frac{dL_{SO}}{dt}\ , \nonumber \\
\frac{dL_{PN}}{dt}  &=& {\frac{2mL}{105\mu^3r^{7}}}\Bigl\{
4E^2\mu^2(-253+281\eta)r^{4} - 12E m\mu^3(324-101\eta)r^{3} \nonumber \\
&&+6\mu\left[8EL^2(21-68\eta) - 7m^2\mu^3(23+6\eta)\right]r^2 \nonumber \\
&&+ 10L^2 m\mu^2(402-203\eta)r - 15L^4(19-72\eta) \Bigr\} \nonumber \\
\frac{dL_{SO}}{dt}  &=& {\frac{8}{15L\mu^2 r^{7}}}{\bf L\cdot}\Bigl\{
2E^2\mu^2 r^{4}{\mbox {\boldmath $\sigma$}} + 12Em\mu^3 r^{3}({\bf S}+{\mbox {\boldmath $\sigma$}}) \nonumber \\
&&+ 3\mu r^{2}\left[EL^2(6{\bf S}+5{\mbox {\boldmath $\sigma$}})
+m^2\mu^3({\bf S}+{\mbox {\boldmath $\sigma$}})\right]
-m\mu^2L^{2}r(11{\bf S}+5{\mbox {\boldmath $\sigma$}})+15L^4{\mbox {\boldmath $\sigma$}} \Bigr\} \ .
\end{eqnarray}

After parameterizing by $\chi$ the radiative change of the conserved quantities and the orbital evolution of the binary system can be further investigated.

\section{Concluding remarks}

In this paper we have analyzed the dynamics and evolution of binaries moving on closed or open orbits. The radial motion can be parameterized in a suitable way for the description of all three types of binary orbits. With the help of their general description we have determined the amount of energy and orbital angular momentum radiated during the encounter of these compact objects. 

These results can be used to investigate the parameter dependence of the hyperbolic encounters of binaries. Depending on the source parameters and the emitted energy these binaries can form a bound system. These highly eccentric burst type sources are among the expected candidates for future observations.

\section{Acknowledgments}
The author thanks Bal\'{a}zs Mik\'{o}czi for useful comments during the preparation of the ma\-nu\-script. This paper was supported by the J\'{a}nos Bolyai Research Scholarship of the Hungarian academy of Sciences. Partial support comes from "NewCompStar", COST Action MP1304.

\appendix
\section*{Appendix: Radiative change of the energy and angular momentum}

With the use of symmetric and trace-free radiative multipole moments Kidder \cite{Kidder} has given the radiative energy and angular momentum losses for point particles. Applying the equations of motion these quantities can be written in terms of dynamical variables characterizing the orbital motion of the binary. For completeness, we summarize here these expressions up to the required accuracy.

The energy loss is given by
\begin{equation}\label{Erad}
\frac{dE}{dt} = - \frac{8m^2 \mu^2}{15r^4} \left( 12v^2 - 11 \dot
r^2 \right) + \frac{dE_{PN}}{dt} + \frac{dE_{SO}}{dt},
\end{equation}
with the PN and SO contributions
\begin{eqnarray}
\frac{dE_{PN}}{dt}  &=& - \frac{2m^2 \mu^2}{105r^4} \biggl[
(785-852\eta)v^4 -160(17-\eta) \frac{m}{r}v^2 + 8(367-15\eta)\frac{m}{r} \dot r^2 \nonumber \\
&& -2(1487-1392\eta)v^2 \dot r^2 + 3(687-620\eta) \dot r^4
+ 16(1-4\eta)\frac{m^2}{r^2} \biggr], \nonumber \\
\frac{dE_{SO}}{dt}  &=& - \frac{8m \mu }{15r^6}  {\bf L_N \cdot}
\left[ {\bf S} \left(27 \dot r^2 -37v^2 -12
\frac{m}{r}\right) + {\mbox {\boldmath $\sigma$}} \left(51 \dot r^2 - 43v^2 +
4\frac{m}{r}\right) \right] \ .
\end{eqnarray}

The angular momentum loss can be written as
\begin{equation}\label{jrad}
\frac{d{\bf J}}{dt} = - \frac{8m \mu}{5r^3} {\bf L_N}
\left( 2v^2 - 3 \dot r^2 + 2 \frac{m}{r} \right) + \frac{d{\bf J}_{PN}}{dt}
+ \frac{d{\bf J}_{SO}}{dt} \ ,
\end{equation}
where
\begin{eqnarray}
\frac{d{\bf J}_{PN}}{dt}  = - \frac{2m \mu}{105r^3}  {\bf L_N}
\biggl\{ && (307-548\eta)v^4 - 6(74-277\eta)v^2 \dot r^2
+ 2(372+197\eta)\frac{m}{r} \dot r^2
\nonumber \\ && \mbox{}
+ 15(19-72\eta) \dot r^4
- 4(58+95\eta) \frac{m}{r}v^2
- 2(745-2\eta)\frac{m^2}{r^2} \biggr\},
\end{eqnarray}
\begin{eqnarray}
\frac{d{\bf J}_{SO}}{dt}  &=& -\frac{4\mu^2}{5r^3} \left\{
\frac{2m}{3r}(\dot r^2 - v^2)({\mbox {\boldmath $\sigma$}} - {\bf S})
- \dot r \frac{m}{3r}{\bf n \times} \left[ {\bf v \times} (7{\bf S} +
5 {\mbox {\boldmath $\sigma$}}) \right]
\right. \nonumber \\ &&
+ \frac{m}{r}{\bf n \times} \left[ ({\bf n \times S})(6 \dot r^2
-\frac{17}{3}v^2 +2\frac{m}{r}) +
({\bf n \times {\mbox {\boldmath $\sigma$}}}) (9\dot r^2 - 8v^2 - \frac{2m}{3r}) \right]
\nonumber \\ &&
+ \dot r {\bf v \times} \left[ ({\bf n \times S})(\frac{29 m}{3 r}
+ 24v^2 -30 \dot r^2) + 5
({\bf n \times {\mbox {\boldmath $\sigma$}}}) (\frac{5m}{3r} +4v^2-5\dot r^2) \right]
\nonumber \\ &&
+ {\bf v \times} \left[ ({\bf v \times S})(18 \dot r^2
-12v^2 -\frac{23m}{3r}) +
({\bf v \times {\mbox {\boldmath $\sigma$}}}) (18\dot r^2 - \frac{35}{3}v^2-9\frac{m}{r}) \right]
\nonumber \\ && \left. + \frac{{\bf L_N}}{\mu^2 r^2} {\bf L_N \cdot} \left[ {\bf S}(30 \dot r^2 -18v^2
-\frac{92m}{3r}) + {\mbox {\boldmath $\sigma$}}(35 \dot r^2 -19v^2 -\frac{71m}{3r}) \right] \right\} \ .
\end{eqnarray}
Note that the leading order and PN contributions to the radiative change of ${\bf J}$ are in the direction of ${\bf L_N}$, while, in general, ${d {\bf J}}_{SO}/dt$ is not. Moreover, as the radiative spin losses appear only at 2PN order \cite{GPV3}, Eq. (\ref{jrad}) represents the change of the orbital angular momentum ${\bf L}$  in our approximation.

\end{document}